\documentclass[aps,prd,twocolumn,showpacs,amsmath,amssymb]{revtex4}

\usepackage{slashed}
\usepackage{graphicx} 
\usepackage{dcolumn} 
\usepackage{bm} 
\usepackage{epsfig}
\usepackage{amsmath}
\usepackage{amssymb}
\usepackage{hyperref}
\usepackage{color}
\usepackage{ulem}

\begin{document}

\newcommand{\bsy}[1]{\mbox{${\boldsymbol #1}$}}
\newcommand{\bvecsy}[1]{\mbox{$\vec{\boldsymbol #1}$}}
\newcommand{\bvec}[1]{\mbox{$\vec{\mathbf #1}$}}
\newcommand{\btensorsy}[1]{\mbox{$\tensor{\boldsymbol #1}$}}
\newcommand{\btensor}[1]{\mbox{$\tensor{\mathbf #1}$}}
\newcommand{\tensorId}{\mbox{$\tensor{\mathbb{\mathbf I}}$}}
\newcommand{\be}{\begin{equation}}
\newcommand{\ee}{\end{equation}}
\newcommand{\bea}{\begin{eqnarray}}
\newcommand{\eea}{\end{eqnarray}}

\title{Finite-temperature electromagnetic responses of relativistic electrons}

\author{C. A. A. de Carvalho$^{1,2}$}

\affiliation{$^{1}$Centro Brasileriro de Pesquisas F\'{\i}sicas - CBPF, Rua Dr. Xavier Sigaud 150, Rio de Janeiro - RJ, 22290-180, Brazil\\
$^{2}$Instituto de F\'{\i}sica, Universidade Federal do Rio de Janeiro - UFRJ, Caixa Postal  68528, Rio de Janeiro - RJ, 21945-972, Brazil}

\date{\today}

\begin{abstract}
We compute the real and imaginary parts of the electric permittivities and magnetic permeabilities for relativistic electrons from quantum electrodynamics at  finite temperature and density. A semiclassical approximation establishes the conditions for neglecting nonlinear terms in the external electromagnetic fields as well as electron-electron interactions.  We obtain both the electric and magnetic responses in a unified manner and relate them to longitudinal and transverse collective plasma oscillations. We demonstrate that such collective modes are thresholds for a metamaterial regime of the electron plasma which exhibits simultaneously negative longitudinal permittities and permeabilities. For nonzero temperatures, we obtain electromagnetic responses given by one-dimensional integrals to be numerically calculated, whereas for zero temperature we find analytic expressions for both their real/dispersive and imaginary/absorptive parts. 
\end{abstract}


\maketitle

\section{Introduction}

We have recently proposed \cite{PRD} that a gas of relativistic electrons, in the long wavelength limit, for weak fields and densities where electron-electron interactions may be neglected, exhibits Drude-like expressions for both electric and magnetic responses. As a consequence, we argued that the relativistic electron gas represents a natural realization of a metamaterial, understood as a system that displays simultaneously negative electric permittivity and magnetic permeability \cite{Veselago, Smith, Pendry}.

As a preliminary test on the accuracy of the approximations, we have compared our predictions with nonrelativistic data from metallic systems \cite{EPL}, whose quasi-free electrons exhibited responses that agreed well with the  theoretical calculations at finite temperature, giving quite good estimates for physical quantities such as plasma frequencies. Clearly, the truly relativistic electron gases found, for instance, in synchrotron beams, are the physical systems required to check theoretical predictions. Since the formalism for the relativistic electron gas \cite{PRD} may be extended to other charged fermions, relativistic plasmas are also an appropriate testing ground. 

Quantum electrodynamics (QED) at finite temperature and density provides a unified  treatment that relates electromagnetic responses to the well-known polarization tensor \cite{AP} with both vacuum and medium contributions. Using QED, we have shown \cite{PRD} that (i) all responses depend on three scalar functions (one for the vacuum; two for the medium), (ii) polarizations depend on both the electric and magnetic fields, just as magnetizations depend on both magnetic and electric fields, and (iii) many aspects of the analysis of electric responses will carry over to the magnetic ones, due to the analogy between permittivities $\epsilon$ and inverse permeabilities $\mu^{-1}$. Furthermore, in the present work, we will show that (iv) electromagnetic responses may be related to collective plasma oscillations and used to calculate the corresponding plasma frequencies.

The present study is devoted to the calculation of the real and imaginary parts of the finite-temperature electromagnetic responses of relativistic electrons. At $T \ne 0$, the problem is reduced to one-dimensional (1D) integrals which involve the Fermi-Dirac occupation numbers for electrons and positrons. At $T=0$, however, as the occupation numbers become step functions, one obtains analytic expressions for the medium contributions. The existence of imaginary parts in longitudinal responses indicates instabilities that may be associated to plasmons decaying into electron-hole pairs or to electron-positron pair creation. Moreover, such results may then be used to calculate the dispersion relations for both longitudinal and transverse plasmons. Although numerical and analytic theoretical results for response functions have already appeared in the literature \cite{previous}, especially in the context of plasma physics, we stress that the present derivation obtains both electric and magnetic responses in a unified way, relates them to physical quantities that may be measured in the previously described experimental scenarios, and connects them to collective plasma oscillations and to the metamaterial regime that occurs at  frequencies where both electric and magnetic responses are simultaneuosly negative. At $T \ne 0$, for instance, we may produce numerical results that will describe finite-temperature effects in hot plasmas.

The article is organized as follows. Section \ref{TF} reviews the theoretical framework of the present article. It consists of a semiclassical expansion that allows us to generate and interpret the various contributions to the electromagnetic responses, essentially establishing the conditions for the use of linear response and non-interacting electrons. Electromagnetic responses are then obtained from the polarization tensor. Section \ref{CPE} relates electromagnetic responses to collective plasmon excitations. Section \ref{RIP} obtains the imaginary and real parts of the response functions at $T \ne 0$ in terms of 1D integrals. Section \ref{anal} exhibits the explicit analytic expressions at $T=0$. Finally, Section \ref{conc} presents our conclusions.

\section {Theoretical framework}
\label{TF}

The QED finite-temperature partition function  $Z=\hbox{Tr} \, e^{-\beta (\hat{H}-\xi\Delta \hat{N}) }$  describes a relativistic electron gas, with fixed $\Delta N= N_e - N_p$ ($N_e$ is  the number of electrons and $N_p$ the number of positrons) at temperature $T=\beta^{-1}$ and chemical potential $ \xi$, interacting with an electromagnetic field $A_\nu$. It may be expressed as a functional integral over gauge and fermion fields  \cite{KG}
\bea
\label{Z}
&& Z=\oint  [dA_\nu] \det (\delta {\cal G}/\delta \Lambda) \, \delta({\cal G}) e^ {-S_{A}[A]} Z_{e}[A],\\
&& Z_e [A] = \oint [i d\psi^{\dagger}][d\psi] e^{- S_{e} [\psi^{\dagger}, \psi, A]}.
\label{Ze}
\eea

The determinant is the Jacobian of the gauge transformation $A_\nu\rightarrow A_\nu - \partial_\nu \Lambda$, while the delta function imposes the gauge condition ${\cal G}[A]=0$, typically ${\cal G}[A]=\partial_\nu A_\nu$. The actions $S_\ast=  \int_0^{\beta} dx_4  \int d^3 x   \,{\cal L}_\ast$ $(\ast = A,\,e)$ have ${\cal{L}}_{A}= -\frac{1}{4} F_{\mu\nu} F_{\mu\nu}$ and ${\cal L}_e=  \bar{\psi}  \Gamma_{A} \psi $, where $\Gamma_{A}=G_{A} ^ {-1} = i \slashed{D} - m - i \xi\gamma_4 $ is the inverse of the electron propagator at finite density in the presence of the gauge field, $\slashed{D} \equiv \gamma . (\partial  - ie A)$ and $\bar{\psi}= \psi^{\dagger} \gamma_4$. We  use $e$ and $m$ for the electron charge and mass, and natural units $\hbar=1, c=1$.  The integral $\oint$ runs over gauge fields obeying $A_\nu(0,\vec{x})=A_\nu(\beta, \vec{x})$ and electron fields obeying $\psi (0,\vec{x})=- \psi(\beta, \vec{x})$ \cite{MtoE}. 

In \eqref{Z}, we write $A_\mu=\bar A _\mu + a_\mu$ where $\bar {A}_\mu$ is a classical solution of the sourceless equation of motion for $A_\mu$, which we identify with an external classical field incident on the electron gas \cite{russo}, and $a_\mu$ is a quantum fluctuation field so that
\be
\label{A1}
{\cal L}_A=-\frac{1}{4} \bar F_{\mu\nu} \bar F_{\mu\nu} - \frac{1}{4} f_{\mu\nu} f_{\mu\nu}.
\ee
We perform the integral over $a_\mu$ before the fermion integration,
\be
\label{int a}
\oint  [da_\mu]  \det (\delta {\cal G}/\delta \Lambda) \, \delta({\cal G}) e^ {-S_{a}[a_\mu, \psi, \bar{\psi}]},
\ee
where $S_a$ is given by the quadratic form
\be
\label{Sa}
S_a=\frac{1}{2} \iint dx dy \,  a_{\mu} (G^\gamma_{\mu\nu})^{-1} a_{\nu} + e \int dx\,  (\bar{\psi} \gamma_\mu \psi) a_{\mu}
\ee
and $\int dx \equiv \int_0^\beta dx_4\int d^3 x$, with  $G^\gamma_{\mu\nu}$ being the photon propagator in the chosen gauge. We take minus the logarithm of the quadratic integral to obtain
\be
\label{4fermi}
S_{e}^{(i)}=-\frac{e^2}{2} \iint dx dy (\bar{\psi} \gamma_\mu \psi)_x G^\gamma_{\mu\nu}(x-y) (\bar{\psi} \gamma_\nu \psi)_y,
\ee
which shows that  the integral over quantum fluctuations of the gauge field leads to electron-electron interactions mediated by the photon propagator. 

The remaining fermionic integral is
\be
\label{fint}
Z^{(sc)}_e [\bar A] = \oint [i d\psi^{\dagger}][d\psi] e^{- S^{(sc)}_{e} [\psi^{\dagger}, \psi, \bar A]},
\ee
with the fermionic semiclassical action given by $S_ e^{(sc)} = S_e + S_e^{(i)}$. Expanding $\exp(-S_e^{(i)})$ to order $\alpha$, the fermion integral reads
\be
\label{4fexp}
Z^{(sc)} _e [\bar A] \cong \oint [i d\psi^{\dagger}][d\psi] e^{- S_{e} [\psi^{\dagger}, \psi, \bar A]} [1- S_e^{(i)}],
\ee 
and by neglecting the interaction term one obtains
\be
\label{3}
Z^{(sc)}_e[\bar A] \cong \det [-\beta \gamma_4 \Gamma_{\bar A}] = \exp \hbox{Tr} \ln [-\beta \gamma_4 \Gamma_{\bar A}]. \\
\ee
This leads to a modified action for the $\bar A_\mu$ field, $S^{(sc)}[\bar A] = S_{\bar A}[\bar A] -Tr \ln [ -\beta \gamma_4 \Gamma_{\bar A}]$, which takes into account the response of the electrons. 

The extremal condition $\delta S^{(sc)}/ \delta \bar A_{\nu} = 0$ gives the equation of motion
\be
\partial_\mu F_{\mu\nu} = - Tr[e \gamma_\nu G_{\bar A}] = J_\nu.
\ee
Splitting $J$ into free $J^F$ (for $ \bar A=0$) and induced $J^I$ currents, we may write
\bea
&& \partial_\mu ( F_{\mu\nu} +  P_{\mu\nu})= J^{F}_\nu, \\
&& -\partial_\mu P_{\mu\nu}= J_\nu^{I}= Tr[e \gamma_\nu G_{\bar A}] - Tr[e \gamma_\nu G_0],
\label{polinear}
\eea
with $G_0$ being the free electron propagator at finite density. $P_{\mu\nu}$ defines the polarization  $\vec P$ ($P_{4j}= iP^j$)  and magnetization $\vec M$ ($P_{ij}= - \epsilon_{ijk} M^k$) vectors.

The electron propagator at finite density may be expanded in the background field, so that $\hbox{Tr} \ln [-\beta \gamma_4 \Gamma_{\bar A}] - \hbox{Tr} \ln [-\beta \gamma_4 G^{-1}_{0}]$ is given as an infinite sum of one-loop graphs, i.e., a fermion loop with an even number (due to Furry's theorem \cite{IZ}) of insertions of the classical field. The first term of the series is
\be
\label{one-loop}
-\frac{e^2}{2} \hbox{Tr} (G_0 \gamma . \bar{A} G_0 \gamma . {\bar A}),
\ee
which may be written as
\be
\label{appol}
\frac{1}{2}\left [\frac{1}{\beta}\sum_n \!\int \!\frac{d^3q}{(2\pi)^3} {\tilde {A}}_\mu (q) {\tilde \Pi}_{\mu\nu}  (q) {\tilde A}_\nu (-q)\right ],
\ee
in terms of the Fourier transform $\tilde A$ of $\bar A$, with ${\tilde \Pi}_{\mu\nu}  (q)$ being the one-loop vacuum polarization tensor \cite{IZ} given by 
\be
\tilde{\Pi}_{\mu\nu} = - \frac {e^2}{\beta} \! \sum_{n=-\infty} ^ {+\infty}\! \int\! \frac {d^3 p}{(2\pi)^3} \hbox{Sp} [\gamma_\mu G_0(p) \gamma_\nu G_0 (p-q)],
\label{pol}
\ee
where the sum is over Matsubara frequencies $p_4= (2n+1) \pi T$ and Sp denotes trace over Dirac matrices.
The next term, with four insertions, is still one-loop but nonlinear in the fields. It depends on $(T,\xi)$ and is typically of order $\alpha^2 E^2/m^4$ or $\alpha^2 B^2/m^4$.

If we consider the first contribution from the $e-e$ interaction, we have to contract the four fermion term in $S_e^{(i)}$ with the electron propagator at finite density in the presence of the external field, which yields a two-loop contribution. When we expand in the external field, the first contribution is quadratic in the field, of order $\alpha^2$, and contributes in the linear response. The next terms in the expansion in the external field are nonlinear $(T, \xi)$-dependent contributions of order $\alpha^2 E^2/m^4$ or $\alpha^2 B^2/m^4$. 

Our calculations neglect nonlinear one-loop contributions, the two-loop contribution to linear response of order $\alpha^2$, and nonlinear ones that also come in with $e-e$ interactions. Although nonlinear terms might bring in interesting effects \cite{refB}, we will restrict our analysis to fields that are not strong enough to invalidate the linear-response approximation. Therefore, we only consider the interaction of independent electrons with weak external fields. 

We note that the expansion of the current in the classical field $\bar A_\nu$ yields an infinite series of one-loop graphs. We only retain the linear term, which amounts to the linear-response approximation. This is related to the well-known RPA approximation \cite{PN, Texts2} of Condensed Matter Physics. Then, Eq. \eqref{polinear} leads to the momentum space equation
\be
i q_\mu \tilde{P} _{\mu\nu}(q)= \tilde{\Pi}_{\nu\sigma}(q) \tilde{A}_\sigma(q).
\label{PPiA}
\ee
whose solution,
\be
\tilde{P}_{\mu\nu} = \frac{\tilde{\Pi}_{\mu\sigma}}{q^2} \tilde F_ {\nu\sigma}- \frac{\tilde{\Pi}_{\nu\sigma}}{q^2} \tilde F_{\mu\sigma},
\label{Pmunu}
\ee
relates polarization and magnetization to the fields $\vec{E}$  ($\tilde F_{4j}= i\tilde E^j$)  and $\vec B$ ($\tilde F_{ij}= \epsilon_{ijk} \tilde B^k$), thus yielding the electric and magnetic susceptibilities and,  ultimately, the electric permittivity and magnetic permeability tensors. 

One may write $\tilde{\Pi}_{\nu\sigma}=\tilde{\Pi}_{\nu\sigma}^{(v)} + \tilde{\Pi}_{\nu\sigma}^{(m)}$ to separate vacuum ($T= \xi = 0$) and medium contributions. The vacuum contribution has the structure
\be
 -\frac{\tilde{\Pi}_{\nu\sigma} ^{(v)}}{ q^2 }= \left (\delta_{\nu\sigma} - \frac{q_\nu q_\sigma}{q^2}\right ) {\cal{C}}(q^2).
\label{vac}
\ee
The medium , however, introduces a preferred reference frame (that of its center of mass). The symmetry of the problem is then reduced to 3D rotations and gauge invariance leading to \cite{AP}

\bea
\label{med1}
&& -\frac{\tilde{\Pi}_{ij} ^{(m)}}{ q^2 }= \left (\delta_{ij} - \frac{q_iq_j}{|\vec{q}|^2}\right ) {\cal{A}} + \delta_{ij} \frac{q_4^2}{|\vec{q}|^2} {\cal{B}},  \\
&& -\frac{\tilde{\Pi}_{44} ^{(m)}}{q^2 } = {\cal B},  \,\,\,\,\,\,\,\,\,  -\frac{\tilde{\Pi}_{4i} ^{(m)}}{q^2 } = - \frac{q_4 q_i}{|\vec{q}|^2} {\cal B},
\label{med2}
\eea
where the three scalar functions ${\cal{A}} (q_4, |\vec{q}|)$, ${\cal B} (q_4, |\vec{q}|)$, and ${\cal{C}}(q^2)$ are determined from the Feynman graph in Eq. \eqref{pol}, the QED polarization tensor at finite temperature, and density. ${\cal{A}}$ and ${\cal{B}}$ are calculated from the $\tilde{\Pi}_{\mu\mu}$ trace and $\tilde{\Pi}_{44}$, once we subtract the vacuum part, i.e,
\be
{\cal{A}}= \frac{-e^{2}}{2 \pi^{3} q^{2}} \mathrm{Re}\! \! \int\! \! \frac{d^3 p}{\omega_p} n_{F} (p) \frac{p.(p+q)}{q^2-2p.q} +\! \left (1- \frac{3q^2}{2|\vec{q}|^2}\right )\! {\cal{B}}, \\
\label{calA}
\ee
\be
{\cal{B}}=\frac{-e^{2}}{2 \pi^{3} q^{2}} \mathrm{Re}\!\! \int\!\! \frac{d^3 p}{\omega_p} n_{F} (p) \frac{p.q - 2p_4(q_4 - p_4)}{q^2-2p.q} ,
\label{calB}
\ee
where $p_4=i\omega_p=i \sqrt{|\vec{p}|^2+ m^2}$ and $n_{F}(p)$ is the Fermi-Dirac occupation number for particles and antiparticles,
\be
n_{F}(p)= \frac{1}{e^{\beta(\omega_p - \xi)} +1} +  \frac{1}{e^{\beta(\omega_p + \xi)} +1}.
\ee

We now introduce $H_{\mu\nu} \equiv F_{\mu\nu}+ P_{\mu\nu}$ which defines $H_{4j} = iD^j$ and $H_{ij}= \epsilon_{ijk} H^k$, with $\vec{D}= \vec{E} + \vec{P}$ and $\vec{H} = \vec{B} - \vec{M}$. The constitutive equations are derived from Eqs. \eqref{vac}, \eqref{med1}, and \eqref{med2}, i.e,
\begin{eqnarray}
\label{Dconst}
&& \tilde D^j=\epsilon^{jk} \tilde E^k + \tau^{jk} \tilde B^k, \\
&& \tilde H^j= \nu^{jk} \tilde B^k + \sigma^{jk} \tilde E^k \, \, ,
\label{Hconst}
\end{eqnarray}
where we have defined $\nu_{jk} \equiv( {\mu^{-1}})_{jk}$ as the inverse of the magnetic permeability tensor. Using $\hat{q}^i\equiv q^i/|\vec{q}|$, we obtain the linear-response tensors
\bea
\label{resp1}
&& \epsilon^{jk}= \epsilon \delta^{jk} + \epsilon' \hat{q}^j \hat{q}^k, \\
\label{resp2}
&& \nu^{jk}= \nu \delta^{jk} + \nu' \hat{q}^j \hat{q}^k, \\
\label{resp3}
&& \tau^{jk}= \tau \epsilon^{jkl} \hat{q}^l, \\
&& \sigma^{jk}= \sigma \epsilon^{jkl} \hat{q}^l.
\label{resp4}
\eea
 Again, $\nu \equiv \mu^{-1}$, $\nu'\equiv \mu'^{-1}$. One should stress that there are contributions  to $(\vec{D}, \vec{H})$ along the directions of the fields $(\vec{E}, \vec{B})$, of the wavevector $\vec{q}$, and of $(\vec{q}\wedge\vec{B}, \vec{q}\wedge\vec{E})$. Also note that bianisotropic crystals satisfy similar relations \cite{Bi}.

The Euclidean space $\tilde{\Pi}_{\mu\nu}$ is a function of $q_4$. In order to obtain Minkowski expressions, we follow the procedure justified in \cite{PRD} and let $q_ 4= \omega_n \rightarrow i\omega - 0^+$. Note that the Euclidean $q^2=q_4^2+|\vec{q}|^2$ goes to the Minkowski $-q^2=-q_0^2+|\vec{q}|^2= -\omega^2 + |\vec{q}|^2$. We then arrive at the Minkowski expressions below, with the asterisk corresponding to $q_ 4= \omega_n \rightarrow i\omega - 0^+$. The permittivities and inverse permeabilities
\bea
\label{responses1}
&& \epsilon=1+(2 - \frac{\omega^2}{q^2} ){\cal C}^\ast+{\cal A}^\ast+( 1- \frac{\omega^2}{|\vec{q}|^2}) {\cal B}^\ast, \\
\label{responses2}
&& \nu=1+(2+ \frac{|\vec{q}|^2}{q^2}){\cal C} ^\ast+ {\cal A}^\ast - 2\frac{\omega^2}{|\vec{q}|^2} {\cal B}^\ast, \\
&& \epsilon'= - \nu'= \frac{|\vec{q}|^2}{q^2}{\cal C}^\ast - {\cal A}^\ast, \\
\label{responses4}
&& \tau= \sigma= \frac{\omega}{|\vec{q}|}( \frac{|\vec{q}|^2}{q^2} {\cal C}^\ast - {\cal B}^\ast),
\eea
are determined by the three scalar functions ${\cal A}^\ast$, ${\cal B}^\ast$, and ${\cal C}^\ast$, where again the asterisk means $q_ 4 \rightarrow i\omega - 0^+$.  ${\cal C}^\ast$ may be obtained at $T=\xi=0$ \cite{IZ}
\be
{\cal C}^\ast = \frac{-e^2}{12 \pi^{2}}\{ \frac{1}{3}+ 2(1+\frac{2m^2}{q^2}) [h \, \mathrm{arccot} (h) -1] \}
\ee
where $h=\sqrt{(4m^2/q^2) -1}$. The renormalization condition is $\alpha=e^2/(4\pi\hbar c)=1/137$, with $e^2=e^2(\omega=0,\vec{q}=\vec{0})$. Note that the vacuum contributions to permittivities and inverse permeabilities are obtained by setting ${\cal A}^\ast= {\cal B}^\ast=0$. 

For further discussions, it is important to diagonalize the tensors appearing in \eqref{resp1}, \eqref{resp2}. For $\epsilon^{jk}$, the eigenvalues $\lambda$ satisfy $\det(\epsilon^{jk} - \lambda \delta^{jk})=0$, leading to $(\epsilon - \lambda)^2 (\epsilon+ \epsilon'- \lambda)=0$. The eigenvector associated to $\epsilon+ \epsilon'$ is $ \hat{q}^k$, so that it is longitudinal, whereas the two eigenvalues $\epsilon$ are transverse to $ \hat{q}^k$. The same occurs for $\nu^{jk}$. Therefore, from \eqref{responses1}, \eqref{responses2} one derives
\bea
\label{long1}
&& \epsilon_L=\epsilon+\epsilon'= 1+{\cal C}^\ast+( 1- \frac{\omega^2}{|\vec{q}|^2}) {\cal B}^\ast, \\
\label{long2}
&& \nu_L=\nu+\nu'=1+2({\cal C} ^\ast+ {\cal A}^\ast - \frac{\omega^2}{|\vec{q}|^2} {\cal B}^\ast).
\eea
Clearly, the linear-response tensors $\tau^{jk}$ and $\sigma^{jk}$ [cf. Eqs. \eqref{resp3} and \eqref{resp4})] are transverse.

\section{Collective plasmon excitations}
\label{CPE}

We return to the partition function \eqref{Z}, normalized by the free fermion result $Z_e[0]$, in Euclidean space. Instead of doing a semiclassical approximation for the electromagnetic fields, we integrate over the electrons to obtain 
\be
\label{normZ}
\frac{Z_e[A]}{Z_e[0]}=\frac{\det[G_A^{-1}]}{\det[G_0^{-1}]}=-\exp \hbox{Tr} \ln [G_{A}G_0^{-1}].
\ee
We now expand in the field $A_\mu$ and keep only the quadratic part
\be
\label{quadZe}
S_e=-\ln \left (\frac{Z_e[A]}{Z_e[0]}\right )= -\frac {1}{2} \sum \!\!\!\!\! \!\!\!\int {\tilde A}_\mu (q) {\tilde \Pi}_{\mu \nu} (q) {\tilde A}_\nu (q),
\ee
where we have used \eqref{pol} and the simplified notation
\be
\sum \!\!\!\!\! \!\!\!\int \equiv \frac {1}{\beta} \!\! \sum_{n=-\infty} ^ {+\infty}\! \!\int\! \frac {d^3 q}{(2\pi)^3}.
\ee
The partition function is then given by a quadratic integral \cite{KG}
\be
\label{quadpart}
\frac{Z[A]}{Z_e[0]}= \oint [dA_\mu] \det[-\partial ^2] \exp ( -\frac {1}{2} \sum \!\!\!\!\! \!\!\!\int {\tilde A}_\mu {\tilde \Gamma}_{\mu \nu} {\tilde A}_\nu),
\ee
where
\be
\label{Gamma}
{\tilde \Gamma}_{\mu \nu}= q^2 \delta_{\mu\nu} - (1- \frac{1}{\lambda}) q_\mu q_\nu - {\tilde \Pi}_{\mu \nu},
\ee
and the determinant comes from the Lorentz gauge condition, with $\lambda$ being a gauge parameter. 

Following \cite{KG}, we introduce the  ${\cal P}_{\mu\nu}$ projectors
\be
\label{proj}
{\cal P}_{\mu\nu}= \delta_{\mu\nu} - \frac{q_\mu q_\nu}{q^2},
\ee 
where $q^2= q_4^2 + |{\vec q} |^2$ and the transverse ${\cal P}^T_{\mu\nu}$
\bea
\label{projT}
&& {\cal P}^T_{ij}= \delta_{ij} - {\hat q}_i {\hat q}_j,\\
&& {\cal P}^T_{44}={\cal P}^T_{4i}=0,
\eea
with ${\hat q}_i = q_i/|{\vec q}|$. We may define a longitudinal projector ${\cal P}^L_{\mu\nu} \equiv {\cal P}_{\mu\nu} - {\cal P}^T_{\mu\nu}$ and the polarization tensor may then be written as 
\be
{\tilde \Pi}_{\mu \nu}={\tilde \Pi}^{(v)}_{\mu \nu} + {\tilde \Pi}^{(m)}_{\mu \nu} = {\cal F} {\cal P}^L_{\mu\nu} + {\cal G} {\cal P}^T_{\mu\nu},
\ee
where 
\bea
\label{F}
&& {\cal F} = -q^2 \left (1+{\cal C}+ {\cal B} + \frac{q_4^2}{|\vec q |^2} {\cal B}\right ), \\
&& {\cal G} = -q^2 \left (1+{\cal C}+ {\cal A} + \frac{q_4^2}{|\vec q |^2} {\cal B}\right ).
\label{G}
\eea
The expression for the quadratic kernel is given by
\be
\label{kernelgamma}
{\tilde \Gamma}_{\mu \nu} = (q^2 - {\cal F}) {\cal P}^L_{\mu\nu} +  (q^2 - {\cal G}) {\cal P}^T_{\mu\nu} + \frac{1}{\lambda} q_\mu q_\nu,
\ee
and its inverse, the photon propagator, reads
\be
{\tilde \Gamma}^{-1}_{\mu \nu} = \frac{ {\cal P}^L_{\mu\nu}}{q^2 - {\cal F}} + \frac{ {\cal P}^T_{\mu\nu}}{q^2 - {\cal G}} + \frac{\lambda}{q^2} \frac{q_\mu q_\nu}{q^2}.
\ee
In Minkowski space  ($q_4 \rightarrow i \omega - 0^+$, $q^2 \rightarrow - q^2$),
\bea
\label{poleE}
&& \frac{1}{q^2 - {\cal F}} \rightarrow \frac{1}{-q^2 \epsilon_L }, \\
&& \frac{1}{q^2 - {\cal G}} \rightarrow \frac{2}{-q^2 [\nu_L + 1]},
\label{poleM}
\eea
which leads to poles in the ${\cal P}^L_{\mu\nu}$ longitudinal and ${\cal P}^T_{\mu\nu}$ transverse propagators whenever, respectively,
\bea
\label{eplasmon} 
&& \epsilon_L (\omega, |{\vec q}|) = 0, \\
&& \nu_L (\omega, |{\vec q}|)= -1.
\label{mplasmon}
\eea

We remark that  \eqref{eplasmon} corresponds to the usual Condensed Matter dispersion relation of longitudinal plasmon collective excitations. Analogously,  Eq. \eqref{mplasmon} yields the dispersion relation of transverse plasmon collective excitations. Indeed, we may write the projectors as
\bea
&& {\cal P}_{\mu\nu} = n^{(1)}_\mu n^{(1)}_\nu + n^{(2)}_\mu n^{(2)}_\nu + n^{(3)}_\mu n^{(3)}_\nu , \\
&& {\cal P}^T_{\mu\nu}=n^{(1)}_\mu n^{(1)}_\nu + n^{(2)}_\mu n^{(2)}_\nu,
\eea 
where $n^{(i)}_\mu= (0, {\hat n}^{(i)})$, ${\hat q} . {\hat n}^{(i)}=0$, $|{\hat n}^{(i)}|=1$, for $i=1,2$,  satisfying ${\hat n}^{(1)}_i{\hat n}^{(1)}_j + {\hat n}^{(2)}_i{\hat n}^{(2)}_j + {\hat q}_i{\hat q}_j = \delta_{ij}$. For $n^{(3)}_\mu$, we find
\be
n^{(3)}_\mu = \left (\frac{-|{\vec q}|}{\sqrt{q^2}}, \frac{q_4 {\hat q}}{\sqrt{q^2}}\right ),
\ee 
if we demand that it must be normalized and orthogonal to $q_\mu$ and $n^{(i)}_\mu , i=1,2$, thus satisfying $n^{(1)}_\mu n^{(1)}_\nu + n^{(2)}_\mu n^{(2)}_\nu + n^{(3)}_\mu n^{(3)}_\nu +( q_\mu q_\nu / q^2)= \delta_{\mu \nu}$. Then
\bea
&& {\cal P}^L_{\mu\nu} = n^{(3)}_\mu n^{(3)}_\nu , \\
&& {\cal P}^T_{\mu\nu}=n^{(1)}_\mu n^{(1)}_\nu + n^{(2)}_\mu n^{(2)}_\nu.
\eea 
A few observations are in order: \\ \\
(i) in Minkowski space, we have
\be
n^{(3)}_\mu = \left (\frac{i |{\vec q}|}{\sqrt{q^2}}, \frac{\omega {\hat q}}{\sqrt{q^2}}\right ),
\ee 
which in the long-wavelength limit becomes $n^{(3)}_\mu= (0, {\hat q})$; \\ \\
(ii) in that limit, we obtained \cite{PRD} Drude expressions for $\epsilon_L= 1- (\omega_e^2/\omega^2)$ and $\nu_L=1- (\omega_m^2/\omega^2)$. Inserting this into \eqref{poleE} and \eqref{poleM}, and using the fact that $\omega_m^2= 2\omega_e^2$, we find $\omega_e^2 - \omega^2$ as the denominator for both longitudinal and transverse propagators.

Finally, the collective plasmon excitations correspond to charge density and current density oscillations. Indeed, the collective field $A^L \equiv n^{(3)}_\mu {\cal P}^L_{\mu\nu} A_\nu = A_\nu n^{(3)}_\nu$, in Euclidean space, is given by
\be
A^L=\frac{ -i {\vec q} . ( -i {\vec q} A_4 + iq_4 \vec{A})}{\sqrt{q^2} |{\vec q}|}  = \frac{- \vec{\nabla} . {\vec E}}{\sqrt{q^2} |{\vec q}|} = \frac{- \rho(q)}{\sqrt{q^2} |{\vec q}|},
\ee
whereas the collective field $A^T_{\mu} \equiv {\cal P}^T_{\mu\nu} A_\nu$  is given by $(0, {\vec A}^T)$, where ${\vec A}^T= A_1 {\hat n}^{(1)} + A_2 {\hat n}^{(2)}$ and $A_i= {\vec A}. {\hat n}^{(i)}$. One then obtains
\be
{\vec A}^T=\frac{ i {\vec q} \wedge ( i {\vec q} \wedge {\vec A})}{|{\vec q}|^2}=\frac{ {\vec \nabla} \wedge {\vec B}}{|{\vec q}|^2} = \frac {{\vec j}(q)}{|{\vec q}|^2}.
\ee
If we use \eqref{F}, \eqref{G}, and \eqref{kernelgamma}, and leave aside a gauge term, the integrand in \eqref{quadZe} may be written, in Minkowski space, as
\be
\label{osc}
\rho(q)\left (\frac{\epsilon_L}{{\vec q}^2}\right )\rho(q) + j_k(q) \left [\frac{(\nu_L+ 1)(1-\frac{\omega^2}{|{\vec q}|^2})}{2{\vec q}^2}\right ]j_k(q),
\ee
where $q=(\omega, {\vec q})$. 

The above expression physically describes the interaction of charge densities induced by the longitudinal component of the fluctuating electric fields and current densities (loops in the plane perpendicular to $\hat q$) induced by the longitudinal component of the fluctuating magnetic fields.

\section{Real and imaginary parts of the responses}
\label{RIP}

We now turn to Eqs. \eqref{calA} and \eqref{calB} which may be integrated over angles and continued to Minkowski space ($q_4 \rightarrow i\omega - 0^+$) to yield
\be
{\cal B}^\ast= \frac{-e^2}{\pi^2 q^2}\! \int_0^\infty \! \frac{dp \, p^2 n_F}{\omega_p} \left (1+ \frac{4\omega_p^2+q^2}{8p|\vec{q}|} f_1 + \frac{\omega_p\omega}{p|\vec{q}|} f_2\right ),
\ee
\be 
{\cal A}^\ast = {\cal D}^\ast + \left (1+ \frac{3q^2}{2|\vec{q}|^2}\right ){\cal B}^\ast,
\ee
where
\be
{\cal D}^\ast = \frac{-e^2}{\pi^2 q^2}\! \int_0^\infty \frac{dp \, p^2 n_F}{\omega_p} \left (1+ \frac{2m^2+q^2}{8p|\vec{q}|} f_1\right ),
\ee
with  $q^2=\omega^2- |\vec{q}|^2$. The functions $f_1$ and $f_2$ are
\be
f_1= \ln \left \{ \frac{(q^2-2p|\vec{q}|)^2 - 4 \omega_p^2\omega^2}{(q^2+2p|\vec{q}|)^2 - 4 \omega_p^2\omega^2}\right \},
\ee
\be
f_2= \mathrm{arctanh} \left (\frac{2\omega\omega_p}{q^2-2|\vec{q}|p}\right )-   \mathrm{arctanh} \left (\frac{2\omega\omega_p}{q^2+2|\vec{q}|p}\right ).
\ee
In terms of the dimensionless variables $x\equiv \omega_p/m$, $y\equiv p/m=\sqrt{x^2-1}$, $a\equiv \omega/2m$,
$b\equiv |\vec{q}|/2m$, and $c^2\equiv a^2 - b^2= q^2/4m^2$, one obtains
\be
\label{f1ab}
f_1= \ln\left \{ \frac{(c^2 -by)^2 - a^2x^2}{(c^2 + by)^2 - a^2x^2}\right \},
\ee
\be
\label{f1f2}
f_2=\ln \left \{ \left |\frac{c^2 - by + ax}{c^2 + by +ax}\right | \right \}- \frac{f_1}{2} .
\ee

\subsection{Imaginary parts}
\label{IP}

From \eqref{f1f2}, $\mathrm{Im} f_2 = - \mathrm{Im} f_1/2$, so that imaginary parts will appear when the argument of the logarithm in $f_1$ becomes negative, i.e., when the product of its numerator $\mathfrak{N}$ times its denominator $\mathfrak{D}$ satisfies,
\be
\mathfrak{N}\mathfrak{D}=[(c^2-by)^2- a^2x^2][(c^2+by)^2-a^2x^2]<0.
\label{biquad}
\ee
The roots of the related biquadratic equation in $x$ ($\ge 0$) are $x_{\pm}= a\pm b\gamma$, where $\gamma^2 \equiv 1-( 1/c^2)$, leading to
\be 
(a-b\gamma)^2<x^2<(a+b\gamma)^2.
\label{cond}
\ee
There are three cases to be considered:
\\

(i) $c^2<0$ ($\gamma>1$; $a<b<b\gamma$), i.e., $-a+b\gamma<x<a+b\gamma$; 

(ii) $0<c^2<1$ ($\gamma$ purely imaginary): \eqref{cond} is never satisfied and $\mathrm{Im}f_1=0$; 

(iii) $c^2>1$ ($\gamma<1$; $a>b>b\gamma$), i.e., $a-b\gamma<x<a+b\gamma$. 
\\

The difference between numerator $\mathfrak{N}$ and denominator $\mathfrak{D}$ is given by $-c^2by$. For case (i), this implies $\mathfrak{N}>0$, $\mathfrak{D}<0$ whereas for case (iii) $\mathfrak{N}<0$, $\mathfrak{D}>0$. As a consequence, for case (ii), $0<c^2<1$, corresponding to the same sign for $\mathfrak{N}$ and $\mathfrak{D}$, we obtain $ \mathrm{Im} {\cal B}^\ast= \mathrm{Im} {\cal D}^\ast=0$. For cases (i) and (iii), we take $\mathrm{Im} f_1 = \varepsilon (\mathfrak{N}) \pi$, where $\varepsilon(r) \equiv \mathrm{sign} (r)$. The choice of sign corresponds to the continuation $q_4 \rightarrow i\omega - 0^+$ [cf. Appendix \ref{APP}]. Therefore
\bea
\label{ImB0}
&& \mathrm{Im} {\cal B}^\ast=\frac{\varepsilon (c^2) e^2}{16 \pi b c^2}\int_{x_l}^{a+b\gamma}\!\!dx\, n_F\, [(x-a)^2-b^2)], \\
&& \mathrm{Im} {\cal D}^\ast= \frac{\varepsilon(c^2) e^2}{32\pi b c^2}\int_{x_l}^{a+b\gamma}\!\! dx\, n_F\, (1+2 c^2).
\label{ImD0}
\eea
For case (i), $c^2<0$, $\mathfrak{N}>0$, and $x_l=-a+b\gamma$ whereas for case (iii), $c^2>1$, $\mathfrak{N}<0$,  and $x_l=a-b\gamma$. It is easy to show that $-a+b\gamma > 1$ in case (i) and $a-b\gamma>1$ in case (iii). Thus, imaginary parts appear for both $ {\cal B}^\ast$ and $ {\cal D}^\ast$ in the regions $c^2<0$ and $c^2>1$ of the $(a,b)$ plane, and they vanish for $0<c^2<1$.

\begin{figure}
\epsfig{file=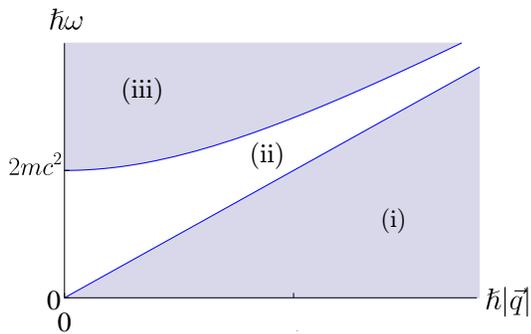,width=.8\columnwidth}
\caption{Regions (i), (ii), (iii) of the $(a,b)=(\hbar\omega, \hbar |{\vec q}|)$ plane.}
\label{fig1}
\end{figure}

Leaving aside the vacuum contribution, the expressions for the longitudinal parts of the electric permittivity and inverse magnetic permeability become
\be
\label{Imelong}
\mathrm{Im} \epsilon_L= - \frac{\varepsilon (c^2) e^2}{16 \pi b^3}\int_{x_l}^{a+b\gamma}\! dx\, n_F\, [(x-a)^2-b^2)],
\ee
\be
\label{Imnlong}
\mathrm{Im} \nu_L= - \frac{\varepsilon (c^2) e^2}{16 \pi b^3}\int_{x_l}^{a+b\gamma}\!\!\!dx\, n_F \left [(x-a)^2+b^2+\frac{b^2}{c^2}\right ].
\ee
Again, they vanish in region (ii) of the $(a,b)$ plane, whiereas in region (i) $x_l=-a+b\gamma$, and in region (iii) $x_l=a-b\gamma$. These various regions are shown in Fig. \ref{fig1}. One should note that the appearance of nonzero imaginary parts may be associated to the creation of electron-hole (lower energies) or electron-positron (higher energies) pairs.

\subsection{Real parts}
\label{RP}

The real parts of ${\cal B}^\ast$ and ${\cal D}^\ast $ may also be reduced to 1D integrals which depend on the  $a=\omega/2m$ and $b=|\vec{q}|/2m$ parameters. We may write them as
\bea
\label{dimB}
&&\mathrm{Re} {\cal B}^\ast= \frac{-e^2}{4\pi^2 c^2}[R+ R_B], \\
&& \mathrm{Re} {\cal D}^\ast = \frac{-e^2}{4\pi^2 c^2}[R + R_D],
\label{dimD}
\eea
where
\bea
&& R= \int_1^\infty dx \,n_F \sqrt{x^2-1}, \\
\label{R}
&& R_B=\frac{1}{4b} \int_1^\infty dx\, n_F [(x^2+c^2) R_1+4ax R_2], \\
\label{dimRB}
&& R_D=\frac{1}{8b} \int_1^\infty dx\, n_F (1+2c^2) R_1.
\label{dimRD}
\eea
with
\be
R_1\equiv \mathrm{Re} f_1 = \ln \left \{ \left | \frac{(c^2-by)^2 -  a^2 x^2}{(c^2+by)^2 -  a^2x^2}\right |\right \},
\ee
\be
R_2\equiv \mathrm{Re} f_2= \frac{1}{2} \ln \left \{ \left | \frac{c^4 - (a x- by)^2}{c^4 - (ax+by)^2}\right |\right \}.
\ee
From the above expressions, one may obtain the electromagnetic responses as functions of $a$, $b$, $T/m$, and $\xi/m$, the latter dependences coming from the Fermi-Dirac distribution function.

\section{Analytic results at $T=0$}
\label{anal}

At zero temperature, the Fermi-Dirac distribution function $n_F(x)$ becomes $\Theta(x_F - x)$ where $x_F \equiv  \xi/m$. We may then analytically perform the integrals for both the imaginary and real parts of the response functions. In the sequel, we will present the details of the calculations and obtain explicit analytic expressions for the imaginary and real parts of both the electric and magnetic response functions as well as their corresponding regions in the $(a,b)$ plane. 

\subsection{Imaginary parts at $T=0$}

Only two of the three cases studied in Section \ref{IP} have nonvanishing imaginary parts. Nevertheless, at zero temperature, as we shall see below, additional restrictions come into play because of the integration limit $x_F$ imposed by the distribution function $\Theta(x_F - x)$.

In case (i) [$c^2<0$, $\gamma>1$ ($a<b<b\gamma$)], the lower integration limit of \eqref{ImB0}, \eqref{ImD0} is $ x_l= -a + b\gamma $ whereas the upper one is $x_u = a+b\gamma$. For a nonvanishing result, we need $x_l = -a +b\gamma < x_F$. This will occur if $b_{-} < b < b_{+}$, where
\be
\label{limits1}
b_{\pm} = \pm \frac{y_F}{2} + \sqrt{\frac{y_F^2}{4} + a(x_F+ a)},
\ee
with $x_F \equiv \epsilon_F /m= \xi /m$ and $y_F = \frac{1}{m}\sqrt{\epsilon_F^2 -m^2}=p_F/m$. Depending on whether $x_u < x_F$ or $x_u>x_F$, results for the imaginary part will differ. For $x_u = a + b\gamma < x_F$, one needs $a< x_F$ and
\bea
\label{condi1}
&& b^2 - y_F b + a(x_F - a) < 0, \\
&& b^2 + y_F b + a(x_F - a) >0.
\label{condi2}
\eea
To satisfy \eqref{condi1}, the argument fo the square-root appearing in the roots of the associated equation has to be positive, so that $0<a<(x_F - 1)/2$ and $(x_F + 1)/2< a< x_F$. Then, \eqref{condi2} will always be satisfied whereas \eqref{condi1} implies $\bar{b}_{-}< b < \bar{b}_{+}$, where
\be
\label{limits2}
\bar{b}_{\pm} = \frac{y_F}{2} \pm \sqrt{\frac{y_F^2}{4} - a(x_F- a)}.
\ee
 As a result, the integrals in \eqref{ImB0} and \eqref{ImD0}, with $n_F = \Theta(x-x_F)$, and the definition
\be 
[f(x)]_{x_l}^{x_u} \equiv f(x_u) - f(x_l),
\ee 
will have the values
\\

A) for $0<a<(x_F - 1)/2$ and $(x_F + 1)/2< a< x_F$, $\bar{b}_{-}< b < \bar{b}_{+}$, $x_l = -a + b\gamma$, $x_u = a+b\gamma$,
\bea
\label{ImB2}
&& \mathrm{Im} {\cal B}^\ast= -\frac{e^2}{48 \pi b c^2} \left[(x-a)^3 - 3 b^2 x\right]_{-a+b\gamma}^{a+b\gamma}, \\ 
&& \mathrm{Im} {\cal D}^\ast= -\frac{e^2}{32 \pi b c^2} \left[(1 + 2c^2) (2a) \right];
\label{ImD2}
\eea

B) for $(x_F - 1)/2<a<(x_F + 1)/2$ and $a>x_F$, ${b}_{-}< b < {b}_{+}$, $x_l = -a + b\gamma$, $x_u = x_F$,
\bea
\label{ImB3}
&& \mathrm{Im} {\cal B}^\ast= -\frac{e^2}{48 \pi b c^2} \left[(x-a)^3 - 3 b^2 x\right]_{-a+b\gamma}^{x_F}, \\ 
&& \mathrm{Im} {\cal D}^\ast= -\frac{e^2}{32 \pi b c^2} [(1 + 2c^2) (x_F + a - b\gamma )].
\label{ImD3}
\eea
Outside regions (A) and (B), for $c^2<0$ the imaginary parts of ${\cal B}^\ast$ and ${\cal D}^\ast$ vanish.

In case (iii) [$c^2>1$, $\gamma<1$ ($a>b>b\gamma$)], the lower integration limit of \eqref{ImB0}, \eqref{ImD0} is $ x_l= a - b\gamma $ whereas the upper one is $x_u = a+b\gamma$. For a nonvanishing result, $x_l = a - b\gamma < x_F$. This will always occur if $a<x_F$ whereas for $a>x_F$, $b'_{-} < b < b'_{+}$, where
\be
\label{limits3}
b'_{\pm} = \pm \frac{y_F}{2} + \sqrt{\frac{y_F^2}{4} - a(x_F- a)}.
\ee
Again, depending on whether $x_u < x_F$ or $x_u>x_F$, results for the imaginary part will differ. For $x_u = a + b\gamma < x_F$, $a< x_F$ and
\bea
\label{condi3}
&& b^2 - y_F b + a(x_F - a) > 0, \\
&& b^2 + y_F b + a(x_F - a) >0.
\label{condi4}
\eea
To satisfy \eqref{condi3}, the argument fo the square-root appearing in the roots of the associated equation has to be positive, so that $0<a<(x_F - 1)/2$ and $(x_F + 1)/2< a< x_F$. Then, \eqref{condi4} will always be satisfied whereas \eqref{condi3} implies $\bar{b}'_{-}< b < \bar{b}'_{+}$, with
\be
\label{limits2}
\bar{b}'_{\pm} = \frac{y_F}{2} \pm \sqrt{\frac{y_F^2}{4} - a(x_F- a)}.
\ee

C) for $0<a<(x_F - 1)/2$ and $(x_F + 1)/2< a< x_F$, $\bar{b}'_{-}< b < \bar{b}'_{+}$, $x_l = a - b\gamma$, $x_u = a+b\gamma$,
\bea
\label{ImB4}
&& \mathrm{Im} {\cal B}^\ast= \frac{e^2}{48 \pi b c^2} \left[(x-a)^3 - 3 b^2 x\right]_{a-b\gamma}^{a+b\gamma}, \\ 
&& \mathrm{Im} {\cal D}^\ast= \frac{e^2}{32 \pi b c^2} \left[(1 + 2c^2) (2b\gamma) \right];
\label{ImD4}
\eea

D) for or $(x_F - 1)/2<a<(x_F + 1)/2$ and $a>x_F$, ${b}'_{-}< b < {b}'_{+}$, $x_l = a - b\gamma$, $x_u = x_F$,
\bea
\label{ImB5}
&& \mathrm{Im} {\cal B}^\ast= \frac{e^2}{48 \pi b c^2} \left[(x-a)^3 - 3 b^2 x\right]_{a-b\gamma}^{x_F}, \\ 
&& \mathrm{Im} {\cal D}^\ast= \frac{e^2}{32 \pi b c^2} [(1 + 2c^2) (x_F -a + b\gamma )].
\label{ImD5}
\eea
Outside regions (C) and (D), for $c^2>1$ the imaginary parts of ${\cal B}^\ast$ and ${\cal D}^\ast$ vanish.

\begin{figure}
\epsfig{file=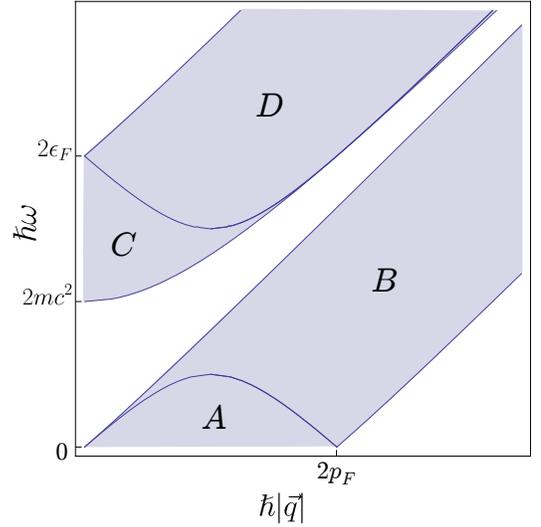,width=.8\columnwidth}
\caption{Regions in the $(a,b)=(\hbar\omega, \hbar |{\vec q}|)$ plane.}
\label{fig2}
\end{figure}

Finally, in case (ii), $0<c^2<1$, one has $\mathrm{Im}{\cal B}^\ast=\mathrm{Im}{\cal D}^\ast=0$. The various regions of the $(a,b)$ plane that correspond to the different expressions discussed above are depicted in Fig. \ref{fig2}.

Before moving on to the real parts of the electromagnetic responses, let us look at the nonrelativistic (NR) limit of cases (A) and (B). At $T=0$, $\epsilon_F=\xi$ and one may define $\epsilon^\prime_F \equiv \epsilon_F - m$. In the  NR-limit, $\epsilon_F \approx 1$, $\epsilon^\prime _F/m << 1$. Furthermore, $\omega \approx |{\vec q}|^2/2m$, so that $a, b <<1$, $\epsilon^\prime_ F/m \approx b^2$, $a^2 \approx b^4$, $a^2 << b^2$, $c^2 < 0$, $ |c^2| << 1$. As a result, $b\gamma \rightarrow 1 + [(a^2+b^4)/2b^2]$ so that  \eqref{ImB2}, \eqref{ImB3} yield \eqref{1c}, \eqref{1b}, respectively (cf. Appendix \ref{APP}). One then recovers the standard nonrelativistic results \cite{Texts2}.

\subsection{Real parts at $T=0$}

At $T=0$, \eqref{R}-\eqref{dimRD} become
\bea
&& R= \int_1^{x_F} dx \, \sqrt{x^2-1},\\
\label{R0}
&& R_B=\frac{1}{4b} \int_1^{x_F} dx\,  [(x^2+c^2) R_1+4ax R_2], \\
\label{dimRB0}
&& R_D=\frac{1}{8b} \int_1^{x_F} dx\,  (1+ 2c^2 ) R_1,
\label{dimRD0}
\eea
where the first integral is simply
\be
R=\frac{1}{2} \left [x_Fy_F - \ln (x_F + y_F)\right ],
\ee 
and the two latter expressions may be integrated by parts. Since $R_1 (1)=R_2 (1)=0$, if we write $R_{B,D}=U_{B,D} + V_{B,D}$, with
\bea
\label{dimUB}
&& U_B=\frac{x_F}{12b} [(x_F^2 + 3 c^2) R_{1F}+ 6ax_F R_{2F}], \\
&& U_D=\frac{x_F}{8b} (1 + 2c^2) R_{1F},
\label{dimUD}
\eea
where
\be
R_{1F} = \ln \left (\left | \frac{(c^2-by_F)^2 -  a^2 x_F^2}{(c^2+by_F)^2 -  a^2x_F^2}\right |\right ),
\ee
\be
R_{2F} = \frac{1}{2} \ln\left ( \left | \frac{c^4 - (a x_F- by_F)^2}{c^4 - (ax_F+by_F)^2}\right |\right ).
\ee
$V_{B,D}$ will involve the derivatives of $R_1$ and $R_2$. We rewrite $\mathrm{Re}f_1$ and $\mathrm{Re}f_2$ in terms of
\bea
\label{L1}
&& L_{a,b}(x) \equiv \ln(|ax+by+c^2|), \\
&& \mathfrak{L}_{\pm a}(x) \equiv L_{\pm a,b}(x) - L_{\pm a,-b}(x),
\eea
to obtain
\bea
\label{log1}
&&\mathrm{Re} f_1= - (\mathfrak{L}_a+\mathfrak{L}_{-a}), \\
&& \mathrm{Re}f_2= -(\mathfrak{L}_a-\mathfrak{L}_{-a})/2.
\label{log2}
\eea
Then, $V_{B,D} = V_{B,D}^+ +V_{B,D}^-$ where
\bea
\label{VBa}
&& V_B^\pm=\frac{1}{12b} \int_1^{x_F} dx\, [x^3 \pm 3ax^2+ 3c^2] \mathfrak{L}_{\pm a}', \\
&& V_D^\pm=\frac{1}{8b} \int_1^{x_F}dx\, (1+2c^2) x \mathfrak{L}_{\pm a}',
\label{VDa}
\eea
with primes denoting derivatives in $x$. Alternatively,
\bea
\label{VBv}
&& V_B^\pm=\frac{1}{6} \int_1^{x_F} \frac{dx}{\sqrt{x^2 - 1}}\, \frac{P_B^\pm (x)}{Q^\pm(x)}, \\
\label{VDv}
&& V_D^\pm=\frac{1}{4} \int_1^{x_F} \frac{dx}{\sqrt{x^2 - 1}} \, \frac{P_D^\pm (x)}{Q^\pm(x)}, 
\eea
with
\bea
\label{PB}
&& P_B^\pm= (x^3 \pm 3ax^2+ 3c^2)  \left (x \pm \frac{a}{c^2}\right ) ,\\
\label{PD}
&& P_D^\pm= \left [(1+2c^2) x \right ] \left (x \pm \frac{a}{c^2}\right ), \\
\label{Q}
&& Q^\pm(x) = x^2 \pm 2ax + d^2,
\eea
where $d^2\equiv (a^2 - b^2\gamma^2)$. By dividing out the polynomials in \eqref{VBv} and \eqref{VDv}  and adding $V_{B,D}^+ + V_{B,D}^-$, one obtains 
\be
V_{B,D}=Y_{B,D}+ Z_{B,D}, 
\ee
where
\bea
&& Y_B= \frac{1}{3} \int_1^{x_F} \frac{dx}{\sqrt{x^2 - 1}} (x^2+ 1-2b^2), \\
&& Y_D= \frac{1}{2} \int_1^{x_F} \frac{dx}{\sqrt{x^2 - 1}}\left (1 + 2c^2\right ),
\eea
which leads to
\bea
&& Y_B=\frac{1 }{6}\left [x_F y_F + \left (3 - 4 b^2\right ) \ln (x_F +y_F)\right], \\
&& Y_D=\frac{1}{2} \left (1+ 2c^2 \right ) \ln (x_F + y_F),
\eea
and $Z_{B,D}$ is given by
\be
\label{ZBD}
Z_{B,D}= C_{B,D} \int_1^{x_F} \frac{dx}{\sqrt{x^2 - 1}} \frac{M_{B,D} \,  x^2 + N_{B,D} }{(x^2 + d^2)^2 - 4a^2 x^2},
\ee
where $C_B= (1/3)$ and $C_D= (1/2)(1 + 2c^2)$. The expressions for $M_{B,D}$ and $N_{B,D} $ depend solely on $a$ and $b$,
\bea
\nonumber
&& M_B = -2a^2(1+4b^2) - [1-2b^2 - 2a^2 (2 - \gamma^2) ] d^2, \\ \nonumber
&& N_B = -d^4 (1-2b^2) , \\ \nonumber
&& M_D = 2 a^2 (1+\gamma^2)  - d^2, \\ \nonumber
&& N_D = - d^4.
\eea
 
Defining $t = \sqrt{(x^2 - 1)/x^2}= y/x$, \eqref{ZBD} reduces to 
\be
Z_{B,D}= C_{B,D}[( M_{B,D} + N_{B,D}) \mathfrak{I}_0  - N_{B,D} \mathfrak{I}_2 ],
\ee
where $\mathfrak{I}_j$ is
\be
\mathfrak{I}_j = \int_0^{t_F} \frac{dt \, t^j}{\mathfrak{C} t^4+\mathfrak{B} t^2+\mathfrak{A}},
\ee
and the coefficients are
\bea
&& \mathfrak{C} = d^4, \\
&& \mathfrak{B} = -2[d^2(d^2 + 1) - 2a^2], \\
&& \mathfrak{A} = (d^2 + 1)^2 -4a^2.
\eea
The above integrals are analytically evaluated in Appendix \ref{App1} and have different expressions depending on whether $\gamma^2 <0$ or $\gamma^2>0$. From \eqref{dimB} and \eqref{dimD}, we obtain
\bea
\label{ReB}
&&\mathrm{Re} {\cal B}^\ast= \frac{-e^2}{4\pi^2 c^2} [U_B+W_B + Z_B] , \\
&& \mathrm{Re} {\cal D}^\ast = \frac{-e^2}{4\pi^2 c^2}[U_D +W_D+Z_D],
\label{ReD}
\eea
where the $W_{B,D} \equiv R + Y_{B,D} $ are
\bea
\label{WB}
&& W_B= \frac{2}{3} [ x_F y_F - b^2 \ln (x_F + y_F) ], \\
&& W_D = \frac{1}{2}[x_F y_F + 2c^2 \ln (x_F + y_F) ],
\label{WD}
\eea
with $Z_{B,D}$ given in terms of $\mathfrak{I}_j$ (cf. Appendix \ref{App1})
\bea
&& Z_B = \frac{1}{3} [(M_B + N_B) \mathfrak{I}_0 - N_B \mathfrak{I}_2], \\
&& Z_D = \frac{1}{2}(1+2c^2) [(M_D + N_D) \mathfrak{I}_0 - N_D \mathfrak{I}_2].
\eea

The results for $\mathrm{Re} {\cal B}^\ast$ and $\mathrm{Re} {\cal D}^\ast$ provide us with analytic expressions for both the electric and magnetic response functions. In particular, the longitudinal responses \eqref{long1} and \eqref{long2} are given by
\bea
\label{longE}
&& \epsilon_L= (1+{\cal C}^\ast) - \frac{c^2}{b^2} {\cal B}^\ast, \\
\label{longB}
&& \nu_L=(1+2{\cal C} ^\ast) +2 {\cal D}^\ast + \frac{c^2}{b^2} {\cal B}^\ast.
\eea
Leaving aside the ${\cal C} ^\ast$ vacuum contribution, $\mathrm{Re}\, \epsilon_L$ and $\mathrm{Re} \,\nu_L$ will follow from \eqref{ReB} and \eqref{ReD}.

We emphasize that the results of the present section yield analytic expressions for the real and imaginary parts of both the electric and magnetic responses of relativistic electrons. As already indicated both in this section and in Appendix \ref{APP}, the $T=0$ nonrelativistic limit of the present theoretical results for the dielectric response coincide with the well-known Lindhardt dieletric function \cite{Lindhard1954}.

\section{Conclusions}
\label{conc}

The present theoretical results for the temperature-dependent electromagnetic responses of the relativistic electron gas provide us with expressions for both imaginary and real parts in terms of 1D integrals. They may be evaluated numerically to yield  functions of $\omega$, $|\vec{q}|$, $\xi$, and $T$. Such functions, in turn, may be used to obtain the dispersion relation for longitudinal plasmons, i.e., the $\omega (|\vec{q}|)$ that satisfies $\epsilon_L =0$, as well as their transverse magnetic analogues for which $\nu_L= -1$. Furthermore, the appearance of nonzero imaginary parts may then be associated to the creation of electron-hole (lower energies) or electron-positron (higher energies) pairs. For $T=0$, the explicit analytic expressions obtained will contain all that information.

The relativistic electron plasma will support longitudinal and transverse plasmons which are related to collective charge density oscillations and collective current density oscillations, respectively. Just as longitudinal plasmons have their dispersion relation given by $\epsilon_L=0$, for transverse ones it is $\nu_L=-1$ that defines the dispersion relation. Clearly then, for frequencies below the longitudinal and transverse plasmon frequencies, both $\epsilon_L$ and $\nu_L$ will be negative, a behavior characteristic of metamaterials. This is typical of relativistic systems, since in the nonrelativistic limit $\nu_L \sim 1$.

\acknowledgments

The author wishes to thank L. E. Oliveira for many fruitful discussions, for a critical reading of the manuscript, and for his warm hospitality at the Institute of Physics of the Universidade Estadual de Campinas, where part of this work was done. Thanks are also due to D. M. Reis and E. Reyes-G\' omez for help with the figures and graphs, and to CNPq and FAPESP for partial financial support.

\appendix

\section{The nonrelativistic limit at $T=0$}
\label{APP}

In order to make contact with standard results \cite{Texts2}, here we outline how to obtain the $T=0$ nonrelativistic expressions  for $\mathrm{Im} {\cal B}^\ast $ from our formulae. We begin with the nonrelativistic expression \cite{PRD} for ${\cal B}$  at $T=0$
\be
{\cal B}=-\frac{e^2}{\pi^2 |{\vec q} |^2} \mathrm{Re}\!\! \int_0^{p_F} \!\! dp \, p^2 \int_{-1}^1 \!\! \frac{d\chi}{\frac{p|{\vec q}|\chi}{m} - \frac{|{\vec q}|^2}{2m} + i q_4},
\ee
where ${p}_F^2=2m \epsilon^\prime_F$, $\epsilon^\prime_F= \xi - m$. Performing the integral over $\chi$, taking the real part of the resulting logarithm, and letting $q_4 \rightarrow i\omega- \eta$, $\eta \rightarrow 0^+$, leads to
\be
\label{NRIm}
{\cal B^\ast}=-\frac{e^2m}{2\pi^2 |{\vec q} |^3} \int_0^{p_F} \!\!\! dp \, p \ln \left (\frac{|{\vec q}|^2 (|{\vec q}| - 2p)^2 - 4m^2\omega^2}{|{\vec q}|^2 (|{\vec q}| + 2p)^2 - 4m^2\omega^2} \right ).
\ee
The argument of the log will be negative for
\bea
&& \frac{m}{|{\vec q}|} (\omega - \epsilon_q) < p <  \frac{m}{|{\vec q}|} (\omega + \epsilon_q) \,\,\, \mathrm{if}\,\,\, \epsilon_q< \omega , \\
&& \frac{m}{|{\vec q}|} (\epsilon_q - \omega) < p <  \frac{m}{|{\vec q}|} (\omega + \epsilon_q) \,\,\, \mathrm{if}\,\,\, \epsilon_q> \omega,
\eea
where $\epsilon_q \equiv |{\vec q}|^2/2m$. We note that in both cases it is the $\mathfrak{N}$ numerator that becomes negative. 

\begin{figure}
\epsfig{file=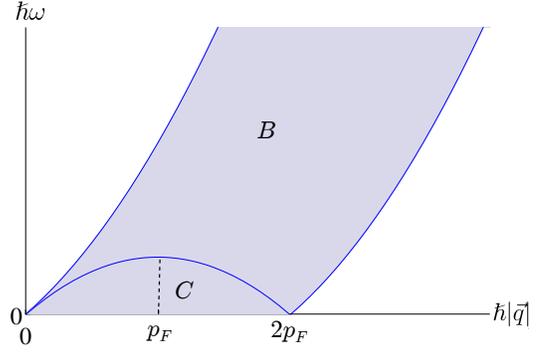,width=.8\columnwidth}
\caption{Regions in the $(\omega,|{\vec q}|)$ plane.}
\label{fig3}
\end{figure}

In order to obtain the imaginary part of the logarithm, we let $q_4 \rightarrow i\omega- \eta$ in
\be
 \ln \left (\frac{|{\vec q}|^2 (|{\vec q}| - 2p)^2 + 4m^2 q_4^2}{|{\vec q}|^2 (|{\vec q}| + 2p)^2 + 4m^2 q_4^2} \right ),
\ee
so \eqref{NRIm} becomes
\be
\label{sign}
\ln \left [ - \left | \mathfrak{N}/\mathfrak{D} \right | + i \eta^\prime (\mathfrak{N} - \mathfrak{D}) \right ],
\ee
where $\eta^\prime \equiv (2ax^2)/\mathfrak{D}^2 \rightarrow 0^+$. Since $\mathrm{sign} (\mathfrak{N} - \mathfrak{D})= \mathrm{sign} (\mathfrak{N})$, the argument of \eqref{sign} will lie in the second quadrant if $\mathrm{sign} (\mathfrak{N})$ is positive and in the third if it is negative. Thus, we choose $[\mathrm{sign} (\mathfrak{N}) \pi]$ for its imaginary part ($-\pi$, in the present case) and therefore $ \mathrm{Im} {\cal B}^\ast $ is given by trivial integrals. \\ \\
{\bf Case 1}: $\omega > \epsilon_q$ \\ \\
{1(a)} if $p_F < m(\omega -\epsilon_q)/|{\vec q}|$, then $\mathrm{Im} {\cal B}^\ast =0$; \\ \\
{1(b)} if $m(\omega -\epsilon_q)/|{\vec q}| < p_F < m(\omega + \epsilon_q)/|{\vec q}|$, the lower integration limit of the imaginary part of \eqref{NRIm} is $m(\omega-\epsilon_q)/|{\vec q}|$ and the upper limit is $p_F$, then
\be
\label{1b}
\mathrm{Im} {\cal B}^\ast= \frac{e^2 m^2 \epsilon^\prime_F}{2 \pi|{\vec q}|^3} \left [1 - \frac{(\omega-\epsilon_q)^2}{4\epsilon^\prime_F \epsilon_q} \right ];
\ee
1(c) if $p_F > m(\omega +\epsilon_q)/|{\vec q}|$, the lower integration limit of the imaginary part of \eqref{NRIm} is $[m(\omega-\epsilon_q)/|{\vec q}|]$ and the upper limit is $[m(\omega + \epsilon_q)/|{\vec q}|]$, then
\be
\label{1c}
\mathrm{Im} {\cal B}^\ast= \frac{e^2 m^2 \omega}{2 \pi|{\vec q}|^3};
\ee
{\bf Case 2}: $\omega < \epsilon_q$ \\ \\
{2(a)} if $p_F < m(\epsilon_q- \omega)/|{\vec q}|$, then $\mathrm{Im} {\cal B}^\ast =0$; \\ \\
{2(b)} if $m(\epsilon_q- \omega)/|{\vec q}| < p_F < m(\epsilon_q +\omega)/|{\vec q}|$, the lower integration limit of the imaginary part of \eqref{NRIm} is $m(\epsilon_q- \omega)/|{\vec q}|$ and the upper limit is $p_F$, then we obtain the same as in \eqref{1b}
\be
\label{2b}
\mathrm{Im} {\cal B}^\ast= \frac{e^2 m^2 \epsilon^\prime_F}{2 \pi|{\vec q}|^3} \left [1 - \frac{(\omega-\epsilon_q)^2}{4\epsilon^\prime_F \epsilon_q} \right ];
\ee
2(c) if $p_F > m(\epsilon_q + \omega)/|{\vec q}|$, the lower integration limit of the imaginary part of \eqref{NRIm} is $m(\epsilon_q - \omega)/|{\vec q}|$ and the upper limit is $m(\epsilon_q + \omega)/|{\vec q}|$, then we obtain the same as in \eqref{1c}
\be
\label{2c}
\mathrm{Im} {\cal B}^\ast= \frac{e^2 m^2 \omega}{2 \pi|{\vec q}|^3}.
\ee
The various regions defined above are plotted in Fig. \ref{fig3}, which is well-known in Condensed Matter texts \cite{Texts2}. As already mentioned previously, the nonrelativistic expressions may also be directly obtained as limits of their relativistic counterparts.
\\
 
\section{The integrals $\mathfrak{I}_j$}
\label{App1}

In this Appendix, we calculate the integrals
\be
\mathfrak{I}_j = \int_0^{t_F} \frac{dt \, t^j}{\mathfrak{C} t^4+\mathfrak{B} t^2+\mathfrak{A}},
\label{Ij}
\ee
with coefficients given by
\bea
&& \mathfrak{C} = d^4, \\
&& \mathfrak{B} = -2[d^2(d^2 + 1) - 2a^2], \\
&& \mathfrak{A} = (d^2 + 1)^2 -4a^2.
\eea

There are two cases to be considered, depending on the roots of the biquadratic equation $\mathfrak{C} t^4+\mathfrak{B} t^2+\mathfrak{A}=0$:

(i) For $\gamma^2 > 0$, the roots are real
\be
\label{tpm}
t_\pm^2= \frac{d^2(d^2+1) - 2a^2 \pm 2ab |\gamma |}{d^4},
\ee
so that the integrals become
\be
\mathfrak{I}_j = \frac{1}{4ab|\gamma |}\left[ t_+^j \int_0^{t_F} \frac{dt}{t^2 - t_+^2} - t_-^j \int_0^{t_F} \frac{dt}{t^2 - t_-^2}\right].
\ee
One may show that $t_\pm^2 > 0$, therefore
\be
\int_0^{t_F} \frac{dt}{t^2 - t_\pm^2}= \frac{1}{2 t_\pm} \ln \left |\frac{t_F - t_\pm}{t_F + t_\pm} \right |,
\ee
where, from \eqref{tpm}, we may write $t_\pm= (y_\pm / |x_\pm |)$, with $x_\pm=a \pm b\gamma$ and $y_\pm = \sqrt{x_\pm^2 - 1}$. Then,
\bea
&& \mathfrak{I}_0 = \frac{1}{4ab|\gamma|} \left [ \frac{1}{2t_+} \ln \left | \frac{t_F - t_+}{t_F+ t_+} \right | - (t_F\rightarrow t_-) \right ], \\
&& \mathfrak{I}_2 = \frac{1}{4ab|\gamma|} \left [ \frac{t_+}{2} \ln \left | \frac{t_F - t_+}{t_F+ t_+} \right | - (t_+ \rightarrow t_-)  \right ]. 
\eea

(ii) For $\gamma^2 < 0$, the roots are complex conjugate ($t_c^2$, $\bar{t}_c^2$), with
\be
\label{tc}
t_c^2= \frac{[d^2(d^2+1) - 2a^2] + i [2ab |\gamma |]}{d^4}.
\ee
We may decompose \eqref{Ij} into partial fractions to obtain $\mathfrak{I}_j= \mathfrak{I}_j^+ - \mathfrak{I}_j^-$ ($t_r \equiv \mathrm{Re}\, t_c; \, t_i \equiv \mathrm{Im}\, t_c$)
\be
\mathfrak{I}_j^\pm = \frac{1}{4d^4 t_r |t_c|^2} \int_0^{t_F} \left[ \frac{dt \, t^j (t \pm 2t_r)}{t^2 \pm 2t_r t + |t_c|^2} \right],
\ee
and write
\bea
&& \mathfrak{I}_0=\frac{1}{8d^4 t_r |t_c |^2} \left \{  \ln \left | \frac{t_F^2+ 2t_r t_F + |t_c |^2}{t_F^2 - 2t_r t_F + |t_c |^2} \right |  +  \right. \nonumber \\ 
&& \left.  \frac{2t_r}{|t_i|} [ \arctan (\frac{t_F+t_r}{|t_i|}  ) + \arctan  (\frac{t_F-t_r}{|t_i|} )  ]  \right \},
\eea
as well as
\bea
&& \mathfrak{I}_2=\frac{1}{8d^4 t_r } \left \{ - \ln \left | \frac{t_F^2+ 2t_r t_F + |t_c |^2}{t_F^2 - 2t_r t_F + |t_c |^2} \right |  +  \right. \nonumber \\ 
&& \left.  \frac{2t_r}{|t_i|} [ \arctan (\frac{t_F+t_r}{|t_i|}  ) + \arctan  (\frac{t_F-t_r}{|t_i|} )  ]  \right \},
\eea
where, from \eqref{tc}, one may show that $t_c = (y_c/x_c)$, with $x_c = a + ib |\gamma|$ and $y_c = \sqrt{x_c^2 - 1}$.


\begin{references}
\bibitem{MtoE} Euclidean and Minkowski coordinates are related by $x_4= i x^0$ and $x_j=x^j$. For gauge fields, $A_4=iA^0$ and $A_j=A^j$. For Dirac matrices, $\gamma_4=\gamma^0$ and $\gamma_j=\gamma^j$. The Minkowski metric is $\eta_{\mu\nu}=(+,-)$.
\end{references}
\end{document}